\documentclass[12pt]{article}
\begin{document}

\hfill {\it \small J. Math. Phys. {\bf 41} 652 (2000)}

\bigskip
\begin{center}
{\Large \bf Hopf algebraic structure of the parabosonic and parafermionic algebras and
 paraparticle generalization of the  Jordan Schwinger map}

\bigskip

\bigskip\bigskip

C. Daskaloyannis${}^\ast$\footnote{daskalo@auth.gr},  K. 
Kanakoglou${}^\ast$ and I. Tsohantjis$^{\dagger}$\\ 

\bigskip
${}^\ast$Physics Department,\\
Aristotle University of Thessaloniki,\\
54006 Thessaloniki, Greece\\

\bigskip
$^{\dagger}$Athens Campus\\
in association with the University of Wales Swansea\\
Faculty of Engineering\\
3, Mac Millan str, 11144 Athens, Greece.

\bigskip
\bigskip

\end{center}

\begin{noindent}
{\bf Abstract}: The aim of this paper is to show that
 there is a Hopf structure of the parabosonic and parafermionic
algebras and this Hopf structure can generate the well known Hopf 
algebraic structure of the Lie algebras, through a realization of 
Lie algebras using the parabosonic (and parafermionic) extension 
of the Jordan Schwinger map. The differences between the Hopf 
algebraic and the graded Hopf superalgebraic structure on the 
parabosonic algebra are discussed. 
\end{noindent}

\bigskip
\bigskip
\bigskip

\vfill



\newpage

\noindent{\bf I. Introduction}\\
Jordan at 1935 proposed a realization of Lie algebras using boson or fermion
 creation - destruction operators.
This realization was rediscovered by Schwinger at 1953, see the discussion of
 the Jordan  - Schwinger map in \cite{BiLo89}[Chap. 5] and in
 \cite{BiLo95}[\S 2.3]. The realization of a Lie algebra by bosons
 corresponds to the symmetric representation of the Lie algebra, while the
  realization by using fermions corresponds to the antisymmetric representation
of the Lie algebra. The case of the u($N$) algebra was presented 
in \cite{Lo70}. The bosons and the fermions are special cases of 
parabosons and parafermions, which are introduced by Green 
\cite{Gr53}. The parabosonic (and correspondingly the 
parafermionic) algebra is a generalization of the usual  bosonic 
(fermionic) algebra leading to generalized alternatives of the 
Bose - Einstein (Fermi- Dirac) statistics or to field theories 
based on paraparticles, all the related bibliography and details 
can be found in\cite{OhKa82}. The $N$ parabosons were used for 
constructing realization  of a sp($2N$) algebra , this 
construction is based on the idea of  using parabosons, rather 
than usual bosons. The same idea can be applied  in the case of 
the so($2N$) algebra but $N$ parafermions are used 
\cite{OhKa82}[\S 3.2]. Biswas and Soni \cite{BiSo88}  used 
systematically parabosons or parafermions for a Jordan -Schwinger 
realization of a u($N$) algebra. In the same paper  realizations 
of the so($2N+1$) or sp($2N$) algebras, using parafermions and 
parabosons, were discussed  in a similar way as in
 \cite{OhKa82,KaTa62,RySu63}. 
 Also  a realization of the g($M/N$) super algebra is proposed
 by using $M$ parafermionic and $N$ parabosonic operators, by extending
the corresponding realizations based on the use of usual fermions and bosons.
Palev\cite{Pa82} has shown that, the
bilinear combinations of the paraoperators yield the superalgebra
gl(n/m) (see also in the same paper the realizations of so(2n+1) and of
osp(1/2m)). Later the same author
 \cite{Pa96} has also proved that the parabosonic and parafermionic algebras
 can be used for constructing realizations of osp($2N+1/2M$) algebras.

The extension of the Jordan  -
Schwinger map  as a method of a realization of every Lie algebra,
using parabosons and parafermions was originally published since 1971 in a local
journal by Palev\cite{Pa71}. This work is not widely known, even we ignored it,
when the first version of this work was printed as a preprint.

The fact that Lie algebras and  superalgebras have a Hopf algebra structure,
constitutes a strong indication that the parabosonic and
parafermionic algebras might possess a  Hopf algebraic structure too.
  If this is true, then the Hopf algebra structure of the Lie algebra
should be consistent with the supposed Hopf algebra structure of the
parabosonic and parafermionic algebra.
Another strong indication of the existence of the Hopf structure in the
parabosonic algebra
can be conjectured by the recent investigation by Macfarlane\cite{McF95},
which  has proved that  the one dimensional Calogero-Vassiliev oscillator
algebra is closely related to the one dimensional  parabosonic algebra.
 In   \cite{TPJ97} and \cite{PT97} a Hopf algebra structure was
 proposed for the one dimensional Calogero-Vassiliev oscillator algebra.
Therefore a natural idea is to transfer the Hopf structure of the
Calogero-Vassiliev algebra to the case of the parabosonic algebra.
In this paper we show that both the parabosonic and parafermionic algebras
admit a Hopf algebra structure.
Also the Jordan-Schwinger map is a Hopf homomorphism from the Lie algebras
 to the parafermionic or parabosonic algebras. That means that the rules of
 the  Hopf algebra structure of the Lie algebra
can be deduced from the rules of the  Hopf algebra structure of 
the parabosonic of parafermionic algebra. Also we show that for an 
Lie algebra of $N$ generators which is realized using  $N$ pairs 
of creation - destruction parabosonic (or parafermionic) 
operators, then the $N$ parabosonic (parafermionic) destruction 
operators are the components of a covariant tensor algebra, while 
the $N$ creation operators are the contravariant components of a 
tensor corresponding to the adjoint representation of the Lie 
algebra.

Ganchev and Palev \cite{GP80} have shown that there is a 
realization of the   Lie superalgebra osp($1,2n$) by using the 
parabosonic algebra. In this realization, the parabosonic 
generators are the odd generators of the  osp($1,2n$). 
 The universal U(osp($1,2n$)) algebra is a 
Z${}_2$ graded associative algebra, which has a usual Hopf 
superalgebra structure \cite{Sch78}. This structure is closely 
related to the Z${}_2$ grading of the osp($1,2n$) algebra.  In 
this paper we show that the Parabosonic algebra is also a Hopf 
algebra, this structure is obtained by adding a generator to the 
parabosonic algebra. Therefore the parabosonic algebra can be 
regarded either as a Hopf superalgebra either as a Hopf algebra. 
The differences of these algebra structures are examined in 
section II. 
 The above cited peculiarity of the parabosonic 
algebra is not valid for the case of parafermionic algebras.

\noindent{\bf II. Parabosonic realization of a Lie Algebra}\\ 
Let consider a finite dimensional Lie algebra  ${\mathcal L} $,
 generated by the generators
$X_1,\,  X_2,\,  \dots X_n$ and the commutation relations:
\begin{equation}
\left[\, X_i,\; X_j \right] = \sum\limits_{k=1}^n c_{ij}^k X_k
\label{eq:LieDef}
\end{equation}
The universal enveloping algebra $U({\mathcal L}) $
is a Hopf algebra with commultiplication,
co-unit and antipode, which are defined by the well known relations:
\begin{equation}
\begin{array}{c}
\triangle(X_i) = X_i \otimes 1 + 1 \otimes X_i\\
\epsilon(X_i)=0 \\
S(X_i) = - X_i
\end{array}
\label{eq:LieHopf}
\end{equation}

Let consider the parabosonic algebra ${\mathcal P}(n)$,
 which is the associative algebra generated
by $n$ generators $b_1,\, b_2\, \ldots , b_n$
 satisfying the trilinear commutation relations:

\begin{equation}
\begin{array}{c}
\left[ \, b_k,  \,  \left\{ b_\ell^\dagger  ,\, b_m \right\} \right]
=  2 \delta_{k \ell} b_m
\\
\left[ \,  b_k,  \,  \left\{ b_\ell^\dagger  ,\, b_m^\dagger   \right\} \right] =
 2 \delta_{k \ell} b_m^\dagger + 2 \delta_{k m} b_\ell ^\dagger
\\
\left[ \,  b_k,  \,  \left\{ b_\ell  ,\, b_m \right\} \right] = 0
%
\end{array}
\label{eq:bos1}
\end{equation}

The generators of the parabosonic algebra define a $u(n)$ algebra by putting
\begin{equation}
N^{[B]}_{\ell  m} = \frac{1}{2} \left\{ b_\ell^\dagger  ,\, b_m \right\}
\label{eq:uNDef}
\end{equation}
The trilinear equations  (\ref{eq:bos1}) imply that the operators $N^{[B]}_{\ell  m}$
 are generators of a
 $u(n)$ algebra, because the following commutation relations are satisfied:
\begin{equation}
\left[\, N^{[B]}_{k\ell}, \; N^{[B]}_{m n} \right] =\delta_{\ell m} N^{[B]}_{k n}
- \delta_{k n} N^{[B]}_{m \ell}
\label{eq:uNCR}
\end{equation}
The linear Casimir operator of the $u(n)$ algebra is defined by:
\begin{equation}
N ^{[B]}= \sum\limits_{i=1}^n N^{[B]}_{ii} =
\frac{1}{2}  \sum\limits_{i=1}^n  \left\{  b_i^\dagger, b_i\right\}
\label{eq:uNCasimir}
\end{equation}
and it satisfies the commutation relations:
\begin{equation}
\left[\, N^{[B]}, b^\dagger_i\right] =  b^\dagger_i,
\qquad  \left[\, N^{[B]}, b_i\right] = - b_i
\label{eq:NCR}
\end{equation}
Starting from the $N^{[B]}$ operator we can define the operator
\begin{equation}
 K= \exp[  i \pi N^{[B]}], \quad  K^{\dagger} = K^{-1} = 
\exp[  - i \pi N^{[B]}] 
\label{eq:DefK} 
\end{equation} 
 The above  commutation relations  imply 
\begin{equation} 
\begin{array}{c}
K K^{\dagger} = K^{\dagger} K = 1\\
\left\{  K,b_i \right\} = \left\{  K,b^\dagger_i \right\}
 =\left\{  K^{\dagger},b_i \right\} = \left\{  K^{\dagger},b^\dagger_i \right\} =0
\end{array}
\label{eq:PropK}
\end{equation}

The parabosonic algebra ${\mathcal P}(n)$ admits a Hopf algebra structure,
with a coproduct $\triangle_p $, co-unit $\epsilon_p$ and antipode $S_p$
given by:
\begin{equation}
\begin{array}{c}
\triangle_p (b_i)  = b_i \otimes  1  + K \otimes b_i\\
\triangle_p (b_i^\dagger)  = b_i^ \dagger \otimes 1 +  K^{\dagger}
  \otimes b^\dagger_i\\
\triangle_p(K)= K\otimes K  \quad
\mbox{and}  \quad  \triangle_p(K^\dagger )= K^\dagger\otimes K^\dagger\\
\epsilon_p(b_i) = \epsilon_p(b^\dagger_i)=0,
\quad  \epsilon_p(K)=\epsilon_p(K^\dagger)=1\\
S_p(b_i)=b_i K^\dagger, \quad
S_p(b^\dagger_i)= b^\dagger_i K, \quad S_p(K)=K^\dagger  ,\quad S_p(K^\dagger)=K
\end{array}
\label{eq:HopfParabose}
\end{equation}
The above defined coproduct $ \triangle_p $ is an algebra 
homomorphism from the space ${\cal P}(n)$ into the tensor product 
 ${\cal P}(n) \otimes {\cal P}(n)$.
  $$ {\cal P}(n) 
\mathop{\longrightarrow}\limits^{\triangle_p}
 {\cal P}(n) \otimes {\cal P}(n) 
$$
 The tensor product  ${\cal P}(n) \otimes {\cal P}(n)$ has the usual tensor algebra
 structure, i.e. there is a product defined as follows:
\begin{equation}
\left( a \otimes b\right)\cdot \left( c \otimes d \right) =
 ac \otimes bd  
\label{eq:TensorProduct}
\end{equation}  
It is not difficult to show that the above definitions do indeed 
satisfy the well known consistency conditions 
\begin{eqnarray}
(id \otimes \Delta) \Delta (a)= (\Delta \otimes id) \Delta (a) \nonumber\\
(id \otimes \epsilon) \Delta (a)= (\epsilon \otimes id) \Delta (a) = a
\nonumber\\
m(id \otimes S) \Delta (a)= m(S \otimes id) \Delta (a) = \epsilon (a)I
\label{def}
\end{eqnarray}

Using the above defined generators a realization of the Lie algebra
 ${\mathcal L}$ can be defined
on the parabosonic algebra ${\mathcal P}(n)$
 by using  a Jordan - Schwinger map\cite{BiLo95}:
$$
{\mathcal L}\ni X_i    \; \mathop{\longrightarrow}\limits^{\mathcal J} \;
 {\mathcal J}(X_i)  \in {\mathcal P}(n)
$$
where
\begin{equation}
 {\mathcal J}(X_i)  = \sum\limits_{k \ell}c_{k i}^\ell N^{[B]}_{k \ell}=
\frac{1}{2}  \sum\limits_{k \ell}c_{k i}^\ell
\left\{ b_k^\dagger  , b_\ell \right\}
\label{eq:LieRealization}
\end{equation}
After trivial calculations one can verify that the generators $ {\mathcal J}(X_i)$
 satisfy
the Lie algebra commutation relations (\ref{eq:LieDef}),
because the structure constants of the Lie algebra
satisfy the Jacobi equality:
$$
c_{ij}^p c_{pk}^m+  c_{jk}^p c_{pi}^m+ c_{ki}^p c_{pj}^m = 0
$$
The Jordan - Schwinger map can be extended to a map from the universal enveloping algebra
$U({\mathcal L})$ to the parabosonic algebra ${\mathcal P}(n)$.
This extension of the  Jordan - Schwinger map, was initially introduced by Palec\cite{Pa71}
since 1971.
Unfortunately, it was published in a local journal, and since its publication,
 it is not widely known.

The existence of the Hopf algebraic structure of the parabosonic algebra  ${\mathcal P}(n)$,
given by equation  (\ref{eq:HopfParabose}),
and the definition of the operators $N^{[B]}_{ij}$, see equation (\ref{eq:uNDef} imply that
$$
\begin{array}{c}
\triangle_p( N^{[B]}_{i j } ) =  N^{[B]}_{i j }\otimes 1 +  1 \otimes N^{[B]}_{i j }\\
\epsilon_p(N^{[B]}_{ij}) =0 \\
S_p(N^{[B]}_{ij}) = -N^{[B]}_{i j }
\end{array}
$$
These relations can be shown by lengthy but trivial algebraic calculations.
 For clarification reasons we reproduce here the proof of the first one
 of these relations.
  $$
 \begin{array}{rl }
 \triangle_p( N^{[B]}_{i j } ) &=
 \frac{1}{2} \triangle_p\left( \left\{b_i^\dagger, b_j\right\}
 \right)
=\frac{1}{2}\left\{ \triangle_p( b_i^\dagger), \triangle_p(b_j) \right\} =\\
&=\frac{1}{2} \left\{  b_i^\dagger\otimes 1 + K^\dagger \otimes  b_i^\dagger,
b_j\otimes 1 + K\otimes b_j \right\}=\\
&=\frac{1}{2}\left(  \left( b_i^\dagger\otimes 1 + K^\dagger \otimes  b_i^\dagger \right)
                     \left( b_j\otimes 1 + K\otimes b_j \right) + \right. \\
& \;\; + \left.      \left( b_j\otimes 1 + K\otimes b_j \right)
                     \left( b_i^\dagger\otimes 1 + K^\dagger \otimes  b_i^\dagger \right) \right)=\\
&=\frac{1}{2} \left( b_i^\dagger b_j \otimes 1 + b_i^\dagger K \otimes  b_j+
                     K^\dagger b_j \otimes  b_i^\dagger+
                      K^\dagger K\otimes b_i^\dagger  b_j  +\right. \\
& \;\;\;\; + \left.   b_j  b_i^\dagger \otimes 1 +  K b_i^\dagger\otimes  b_j+
                      b_j K^\dagger  \otimes  b_i^\dagger+
                      K K^\dagger\otimes   b_j b_i^\dagger
                      \right)=\\
&= \frac{1}{2} \left( \left\{ b_i^\dagger, b_j \right\}\otimes 1 +
                 \left\{ b_i^\dagger, K \right\}\otimes   b_j+
                 \left\{  K^\dagger,  b_j \right\} \otimes   b_i^\dagger+
                 1\otimes \left\{   b_i^\dagger,  b_j  \right\}
                 \right)=\\
&= N^{[B]}_{i j }\otimes 1 +  1 \otimes N^{[B]}_{i j }
\end{array}
$$
By using the parabosonic realization
  (\ref{eq:LieRealization}) of the Lie algebra ${\cal L}$, the
familiar Hopf algebra relations are satisfied:

$$
\begin{array}{c}
\triangle_p\left(
{\cal J}(X_i)\right) = {\cal J}(X_i) \otimes 1 + 1 \otimes {\cal J}(X_i) =
\left({\cal J}\otimes {\cal J}\right)\circ \Delta(X_i)   \\
\epsilon_p\left({\cal J}(X_i)\right)=0=\epsilon(X_i) \\
S_p\left({\cal J}(X_i)\right) = - {\cal J}(X_i)= {\cal J}\left( S(X_i) \right)
\end{array}
$$

Therefore we have shown that the trilinear parabosonic definition (\ref{eq:bos1}) and
the "deformed" -like Hopf structure  (\ref{eq:HopfParabose}) of the parabosonic algebra
imply the Hopf structure of the u(n) algebra.

A direct implication of the above relations is that,
the Hopf algebraic structure of the enveloping algebra  $U({\cal L})$
can be deduced from the Hopf algebraic structure of the parabosonic algebra
${\mathcal P}(n) $.

The following diagrams are commutative:

\begin{equation}
\begin{array}{ccc}
U({\mathcal L})
  & \mathop{\longrightarrow}\limits^\triangle &  U({\mathcal L})\otimes U({\mathcal L}) \\
\\
{}^{\mathcal J}\Big\downarrow         &                                                                  &
\Big\downarrow {}^{ {\mathcal J}\otimes{\mathcal J}}\\
\\
{\mathcal P}(n)  &  \mathop{\longrightarrow}\limits^{\triangle_p} &
 {\mathcal P}(n)\otimes {\mathcal P}(n)
 \end{array}
\label{eq:DeltaDiagram}
\end{equation}

\begin{equation}
\begin{array}{ccc}
U({\mathcal L})   & \mathop{\longrightarrow}\limits^\epsilon    &  C \\
\\
{}^{\mathcal J}\Big\downarrow         &                                                                  &
\Big\downarrow  {}^{ \rm Id }\\
\\
{\mathcal P}(n)  & \mathop{\longrightarrow}\limits^{\epsilon_p}&  C
\end{array}
\label{eq:EpsilonDiagram}
\end{equation}

and
\begin{equation}
\begin{array}{ccc}
U({\mathcal L})   & \mathop{\longrightarrow}\limits^S   &  U({\mathcal L})  \\
\\
{}^{ \mathcal J }\Big\downarrow         &                                                                  &
\Big\downarrow  {}^{ \mathcal J }\\
\\
{\mathcal P}(n)  & \mathop{\longrightarrow}\limits^{S_p}&  {\mathcal P}(n)\end{array}
\label{eq:AntipodeDiagram}
\end{equation}
The above diagrams prove that,  the Jordan map ${\mathcal J}$ is a
Hopf algebra homomorphism\cite{Abe77}

$$
\begin{array}{c}
\triangle_p \circ  {\mathcal J}  =  \left( {\mathcal J}\otimes{\mathcal J} \right)
\circ \triangle \\
\epsilon_p \circ  {\mathcal J} = \epsilon\\
S_p \circ  {\mathcal J}  = {\mathcal J} \circ  S
\end{array}
$$

The realization of any Lie algebra by using boson or fermions operators
was initially discovered by Jordan and
later was rediscovered by Schwinger of the $su(2)$ case,
see \cite{BiLo95}[\S 2.4]. In this construction
the formula (\ref{eq:LieRealization}) is used, but the $b_i$'s correspond to usual bosons.
In this paper,  we have extended the notion of the "Jordan map"
from the boson (or fermion ) case to the paraboson case, i.e. and we have proved that:

\medskip
\begin{noindent}
{\bf Proposition 1}:
{
\em The extended Jordan map, which is defined by (\ref{eq:LieRealization}),
 is a Hopf algebra homomorphism
from the Hopf algebra $U({\mathcal L})$ into the parabosonic Hopf algebra ${\mathcal P}(n)$
$$ U({\mathcal L}) \; \mathop{\longrightarrow}\limits^{\mathcal J}  \; {\mathcal P}(n)$$
}
\end{noindent}
\medskip

Another trivial result is:

\medskip
\begin{noindent}
{\bf Proposition 2}:
{
\em
The set of n parabosonic destruction (or creation) operators
 $\left\{b_i\right\}, \; i=1,\ldots,n$
(or $\left\{b_i^\dagger \right\}, \; i=1,\ldots,n$ )
 are adjoint tensor covariant (correspondingly contravariant) operators of the
Lie algebra ${\mathcal L}$
}
\end{noindent}

That is true because the definition  (\ref{eq:LieRealization})
of the parabosonic realization of the Lie algebra generators and
 the trilinear commutation relations
for the parabosons (\ref{eq:bos1}) imply the relations:
\begin{equation}
\begin{array}{c}
\left[ {\mathcal J}(X_i),  b_j \right] = \sum\limits_{k=1}^n c_{j i}^k b_k \\
 \left[ {\mathcal J}(X_i),  b_j ^\dagger \right] =- \sum\limits_{k=1}^n c_{i k}^j b_k^\dagger
\end{array}
\label{eq:Tensor}
\end{equation}
and the structure constants $c_{ij}^k$ are the matrix elements of the adjoint
representation of the
Lie generators $X_i$.

It should be noted that a Hopf superalgebra structure can be 
obtained by using the results of ref. \cite{GP80}, where the 
parabosonic algebra ${\cal P}(n)$ is formulated as a graded 
Z${}_2$ superalgebra with the parabosonic annihilation and 
creation operators being graded as odd elements of the osp($1,2n$) 
Lie superalgebra. In this structure the parabosonic algebra
 ${\cal P}(n)$ is mapped in the universal enveloping algebra  U(osp($1,2n$)).
 The  graded algebra  U(osp($1,2n$)) is a Hopf superalgebra \cite{Sch78}, 
 that means that there is a superalgebra coproduct 
 $\triangle_{gr}$, a counit $\epsilon$ and a superalgebra antipode $S_{gr}$, 
 which have the usual Hopf algebra properties (\ref{def}).
 These operators can be defined on the parabosonic algebra ${\cal P}(n)$.
  The coproduct  $\triangle_{gr}$ is a mapping from the algebra  ${\cal P}(n)$
  into the graded tensor superalgebra $ {\cal P}(n)\otimes_{gr} {\cal P}(n) $.
  The algebra  $ {\cal P}(n)\otimes_{gr} {\cal P}(n) $ is the usual tensor algebra
  with a graded multiplication, namely:
  \begin{equation}
\left( a \otimes b\right)\cdot \left( c \otimes d \right) = 
 (-1) ^{ deg(b) deg (c) }ac \otimes bd  
\label{eq:GradedTensorProduct} 
\end{equation}  
and 
\begin{equation}  
\begin{array}{c}
\triangle_{gr}(a) = a \otimes 1 + 1 \otimes a \\
 \epsilon(a)=0, \quad \epsilon(1) =1 \\
 S_{gr}(a)= -a, \quad S_{gr}(a b) = (-1)^{deg(a) deg(b)} S_{gr}(b) S_{gr}(a) 
\end{array}
\end{equation}

 Therefore the parabosonic ${\cal P}(n)$ is a Hopf superalgebra \cite{Sch78}, 
 also it is a Hopf algebra as we have shown in this section by adding an additional
 generator $K$ which is defined by equation (\ref{eq:DefK}).

\medskip

\noindent{\bf III. Parafermionic realization of a Lie algebra}\\ 
Another similar construction can be achieved by considering the 
parafermionic algebra ${\mathcal F}(n)$, which is generated by the 
elements $f_1,f_2,\ldots , f_n$, which satisfy the trilinear 
commutation relations: 
\begin{equation}
\begin{array}{c}
\left[ \, f_k,  \,  \left[ f_\ell^\dagger  ,\, f_m \right] \right] =  2 \delta_{k \ell} f_m
\\
\left[ \,  f_k,  \,  \left[ f_\ell^\dagger  ,\, f_m^\dagger   \right] \right] =
 2 \delta_{k \ell} f_m^\dagger - 2 \delta_{k m} f_\ell ^\dagger
\\
\left[ \,  f_k,  \,  \left[ f_\ell  ,\, f_m \right] \right] = 0
\end{array}
\label{eq:fer1}
\end{equation}
This algebra has a simple Hopf algebraic structure given by:
\begin{equation}
\begin{array}{c}
\triangle_f (f_i)  = f_i \otimes  1  + 1 \otimes f_i\\
\triangle_f (f_i^\dagger)  = f_i^ \dagger \otimes 1 +  1  \otimes f^\dagger_i\\
\epsilon_f(f_i) = \epsilon(f^\dagger_i)=0\\
S_f(f_i)= - f_i , \quad
S_f(f^\dagger_i)= - f^\dagger_i
\end{array}
\label{eq:HopfParaFermi}
\end{equation}
and can be easily checked that the consistency relations (\ref{def}) are satisfied.
$u(n)$ generators  can be defined,  similarly as it was given    in equation
(\ref{eq:uNDef}):
\begin{equation}
N^{[F]}_{\ell  m} = \frac{1}{2} \left[ f_\ell^\dagger  ,\, f_m \right],
\qquad N^{[F]}=\sum\limits_{i=1}^n N^{[F]}_{ii}
\label{eq:uNDefF}
\end{equation}
the Lie algebra can be realized by using a Jordan - Schwinger map:
\begin{equation}
 {\mathcal J}(X_i)  = \sum\limits_{k \ell}c_{k i}^\ell N^{[F]}_{k \ell}=
\frac{1}{2}  \sum\limits_{k \ell}c_{k i}^\ell
\left[ f_k^\dagger  , f_\ell \right]
\label{eq:LieRealizationF}
\end{equation}
The generalization of the Propositions 1 and 2 in the case of parafermions is straightforward:

\medskip
\begin{noindent}
{\bf Proposition 1a}:
{
\em The extended Jordan map, which is defined by (\ref{eq:LieRealizationF}),
 is a Hopf algebra homomorphism
from the Hopf algebra $U({\mathcal L})$ into the parafermionic Hopf algebra ${\mathcal F}(n)$
$$ U({\mathcal L}) \; \mathop{\longrightarrow}\limits^{\mathcal J}  \; {\mathcal F}(n)$$
}
\end{noindent}
\medskip

and

\medskip
\begin{noindent}
{\bf Proposition 2a}:
{
\em
The set of n parafermionic destruction
 (or creation) operators $\left\{f_i\right\}, \; i=1,\ldots,n$
(or $\left\{f_i^\dagger \right\}, \; i=1,\ldots,n$ )
 are adjoint tensor covariant (correspondingly contravariant) operators of the
Lie algebra ${\mathcal L}$
}
\end{noindent}

\medskip

\noindent{\bf IV. Discussion}\\
The realizations of the Lie algebras, by using a generalized Jordan map
  seems to be  a useful tool.
 It is well known the importance
 of the Jordan -Schwinger map \cite{BiLo89} in the study
  of the representations of the Lie
Algebras.
 The proposed extension of this fundamental map in the case of the parabosonic and
parafermionic algebra, opens several problems to be solved:
\begin{enumerate}
\item  {\em The relation
of the parabosonic and parafermionic realizations to the known representations
  of the Lie algebras.}\\
 It is well known that the usual bosonic
(or  fermionic) Jordan map leads to the symmetric (antisymmetric) unitary representations
of the Lie algebras. The representations of the parabosonic
and parafermionic algebras are characterized by
a positive integer $p$ and a vacuum state $|0 >$:
$$
b_i b_j^{\dagger} |0> = p \delta_{ij} |0>
$$
Starting from this representation one can construct a representation of the Lie algebra.
The connection of this representation to the known representations is not yet known.
The proposed parabosonic extension of the Jordan - Schwinger map uses the adjoint
representation of the Lie algebra. A more general extension of this construction
can be defined by using the other representations of the Lie algebra.

\item {\em The construction of the dual Hopf algebra corresponding to the Hopf algebra
 of parafermions and parabosons.}
 The dual Hopf algebra of the universal enveloping algebra $U({\mathcal L})$
 is the set of the smooth
 real functions C${}^\infty (G) $ defined on the local Lie group
 corresponding to the Lie algebra ${\mathcal L}$.
The structure of the dual Hopf algebra of the parabosonic or the
parafermionic algebra is not known.

\item {\em The q-deformed versions of parabosonic or the parafermionic algebras}
Another open problem is the q-deformed extension of the Jordan-Schwinger map. It is
well known that the ordinary one dimensional q-deformed bosons
leads to the Jordan -Schwinger map in the special case of the su$_q(2)$,
and analogous constructions are known
for the u${}_q(N)$ algebras. As far as we know, there is not any generalization of the
Jordan - Schwinger map for the q-deformed versions of the parabosonic of parafermionic
algebra, while  descriptions of the  osp${}_q(1/m)$\cite{PaVdJ95}
and  osp${}_q(2n+1/m)$ have been recently studied \cite{Pa98} by using
$q$-deformed generalizations of parabosonic and parafermionic algebras.
A very interesting problem seems to be  the investigation of the $q$-deformed
 generalizations of the parabosonic or parafermionic algebras and their relation with the
Hopf algebraic structure of the quantum groups.

\end{enumerate}

Beyond the importance of the study of representations of the generalized
 Jordan - Schwinger map, we have shown that this map is a Hopf algebra homomorphism.
 This fact implies
that, the Hopf algebraic structure of the universal enveloping algebra of a Lie algebra is
generated by the Hopf algebraic structure of the parabosonic and parafermionic algebras.
From this point of view, the Hopf structure of parabosonic and parafermionic algebras is
more fundamental, than the Hopf structure of Lie algebras.
The extension of these ideas in the case of superalgebras is under investigation.

We must also notify, that Greenberg\cite{Gr64}
conjectured, that the quarks are parafermions of order $p=3$.
The basic symmetry group of the naive quark theory is the SU$(3)$, therefore
the Greenberg's assumption is related to the proposed realization of the
su$(3)$ algebra by using eight parafermions of order   $p=3$.

\noindent {\em Acknowledgements} The important paper \cite{Pa71}, 
on the generalization of the Jordan - Schwinger map, was brought 
in our attention by Prof. T. D. Palev, after the appearance of the 
first preprint version of this paper. We  thank him for very 
useful comments on this paper. 
 Also we acknowledge the useful comments of the referee on the 
 Hopf superalgebraic structure of the parabosonic algebras.
 
This paper was performed with the  support of the Greek 
Secretariat of Research and Technology under contract $\Pi E N E 
\Delta$ 1981/95. 

\bigskip\bigskip



\begin{thebibliography}{99}


\bibitem{BiLo89}
Biedenharn L.C. and  Louck J. D.
{\em Angular Momentum in Quantum
Physics: Theory and Applications,
Encyclopedia of Mathematics and Its Applications {\bf 8}},
Cambridge Univ. Press (UK) (1989)


\bibitem{BiLo95}
Biedenharn L.C. and Lohe M.A.
{\em Quantum Group Theory and q-Tensor Algebras},
World Scientific, Singapore (1995)


\bibitem{Lo70}
Louck J. D.
{\em Recent Progress toward a theory of tensor operators in the unitary groups},
Am. J. Phys. {\bf 38} (1970) 3


\bibitem{Gr53}
Green H. S.
{\em  A Generalized Method of Field Quantization}
Phys, Rev. {\bf 90} (1953) 270


\bibitem{OhKa82}
Ohnuki Y, and Kamefuchi S.
{\em Quantum Field Theory and Parastatistics},
Univ. of Tokyo Press, Springer NY (1982)


\bibitem{BiSo88}
Biswas S. N.  and Soni S. K.
{\em Supersymmetry, parastatistics and operator realizations of a Lie algebra},
J. Math. Phys. {\bf 29} (1988) 16.

\bibitem{KaTa62}
Kamefuchi S. and Takahashi Y.
 {\em A Generalization of Field Quantization and Statistics}, 
Nucl. Phys. {\bf 36} (1962) 177.

\bibitem{RySu63}
Ryan C. and Sudarshan E. C. G.
 {\em Representations of Parafermi Rings},
Nucl. Phys. {\bf 47} (1963) 207.  


\bibitem{Pa82}
 Palev, T.D.
 {\em Para-Bose and para-Fermi operators
      as generators of orthosymplectic Lie superalgebras},
Journ. Math. Phys. {\bf 23} (1982) 1100  (see also Preprint ICTP 
IC/79/167 (1979)). 

\bibitem{Pa96}
Palev, T.D.
{\em A description of the superalgebra osp(2 n + 1/2 m) via Green generators},
J. Phys. A:Math. Gen. {\bf29} (1996) L171


\bibitem{Pa71}
Palev, T. D.
{\em  Second-order realizations
      of Lie algebras with parafields operators},
Comptes Rendus de
l'Academie Bulgare des Sciences,  {\bf 24}  (1971)  565

\bibitem{McF95}
Macfarlane A. J.
{\em Algebraic structure of parabose Fock space. I. The Green's ansatz revisited},
J. Math. Phys. {\bf 35} (1995) 1054




\bibitem{TPJ97}
Tsohantjis I., Paolucci A. and P. D. Jarvis
{\em On boson algebras as Hopf algebras},
J. Phys. A:Math. Gen. {\bf 30} (1997) 4075


\bibitem{PT97}
Paolucci A. and Tsohantjis I.
{\em Hopf-type deformed oscillators, their quantum double and a
q-deformed Calogero-Vasiliev algebra},
Phys. Lett. A{\bf 234} (1997) 27

\bibitem{GP80}
Ganchev A. Ch. and Palev T. D. {\em A Lie superalgebraic 
interpretation of the para-Bose statistics},
 J. Math. Phys. {\bf 21} (1980) 757 

\bibitem{Sch78}
Scheunert M. {\em The Theory of Lie Superalgebras}, Lecture Notes 
in Mathematics vol. 716, ed. Springer NY (1978). 

\bibitem{Abe77}
Abe E.
{\em Hopf algebras},
Cambridge University Press, Cambridge(UK) (1977)

\bibitem{PaVdJ95}
Palev T.D. and Van der Jeugt J.
{\em
The quantum  superalgebra
U${}_q$[osp$(1/N)$: deformed parabose operators and root of unity representations},
 J. Phys. A:Math. Gen. {\bf 28} (1995) 2605


\bibitem{Pa98}
Palev, T.D.
{\em A $q$-deformation of the parastatistics and an alternative to
the Chevalley description of U${}_q$(osp$(2n+1/m)$)},
Comm. Math. Phys. {\bf 196} (1998) 429




\bibitem{Gr64}
Greenberg O. W.
{\em Spin and Unitary-Spin Independence in a Paraquark Model of
Baryons and Mesons},
Phys. Rev. Lett. {\bf 13} (1964) 598



\end{thebibliography}
\end{document}